\pdfoutput=1
\documentclass[runningheads]{springer_template/llncs}

\usepackage{times}
\usepackage{latexsym}
\usepackage[T1]{fontenc}
\usepackage[utf8]{inputenc}
\usepackage{microtype}
\usepackage{inconsolata}
\usepackage{graphicx}
\usepackage{amsmath,amssymb}
\usepackage{xcolor} 
\usepackage{caption}
\usepackage{subcaption}
\usepackage{adjustbox}
\usepackage{breqn}

\newcommand{\encoder}[0]{\mathcal{E}_{vae}}
\newcommand{\decoder}[0]{\mathcal{D}_{vae}}
\newcommand{\generator}[0]{\text{G}}
\newcommand{\discriminator}[0]{\mathcal{D}_{gan}}

\title{SupResDiffGAN a new approach for the Super-Resolution task}

\author{
Dawid Kopeć\inst{1} \orcidID{0009-0000-1765-5810} \and
Wojciech Kozłowski\inst{1} \orcidID{0009-0009-1532-4415} \and
Maciej Wizerkaniuk\inst{1} \orcidID{0009-0001-7337-6418} \and
Dawid Krutul\inst{1} \orcidID{0009-0006-7501-4353} \and \\
Jan Kocoń\inst{1} \orcidID{0000-0002-7665-6896} \and
Maciej Zięba\inst{1} \orcidID{0000-0003-4217-7712}
}
\authorrunning{D. Kopeć et al.}
\institute{
WUST, Wybrzeże Stanisława Wyspiańskiego 27, 50-370 Wrocław, Poland 
\email{\{wojciech.kozlowski, jan.kocon, maciej.zieba\}@pwr.edu.pl}\\
}

\begin{document}
\maketitle
\begin{abstract}
In this work, we present SupResDiffGAN, a novel hybrid architecture that combines the strengths of Generative Adversarial Networks (GANs) and diffusion models for super-resolution tasks. By leveraging latent space representations and reducing the number of diffusion steps, SupResDiffGAN achieves significantly faster inference times than other diffusion-based super-resolution models while maintaining competitive perceptual quality. To prevent discriminator overfitting, we propose adaptive noise corruption, ensuring a stable balance between the generator and the discriminator during training. Extensive experiments on benchmark datasets show that our approach outperforms traditional diffusion models such as SR3 and I$^2$SB in efficiency and image quality. This work bridges the performance gap between diffusion- and GAN-based methods, laying the foundation for real-time applications of diffusion models in high-resolution image generation.
\end{abstract}

\section{Introduction}

\begin{figure*}[t]
    \centering
    
    \begin{minipage}{0.28\textwidth}
        \centering
        \includegraphics[width=\linewidth]{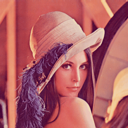}
    \end{minipage}
    \begin{minipage}{0.28\textwidth}
        \centering
        \includegraphics[width=\linewidth]{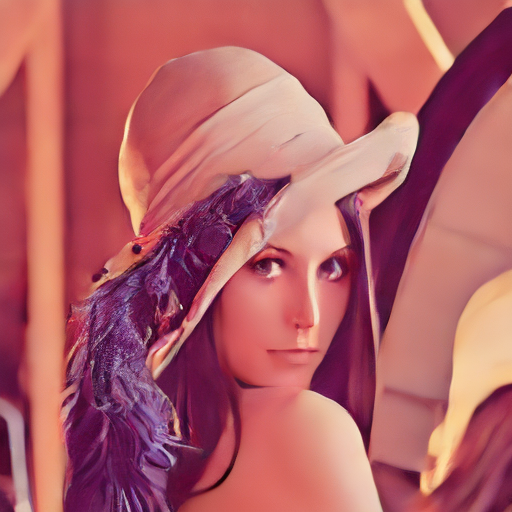}
    \end{minipage}
    \begin{minipage}{0.28\textwidth}
        \centering
        \includegraphics[width=\linewidth]{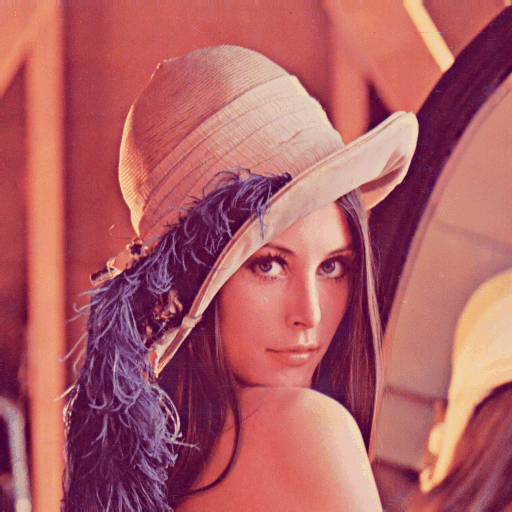}
    \end{minipage}
    
    \begin{minipage}{0.28\textwidth}
        \centering
        \includegraphics[width=\linewidth]{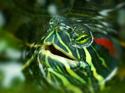}
        \caption*{LR input}
    \end{minipage}
    \begin{minipage}{0.28\textwidth}
        \centering
        \includegraphics[width=\linewidth]{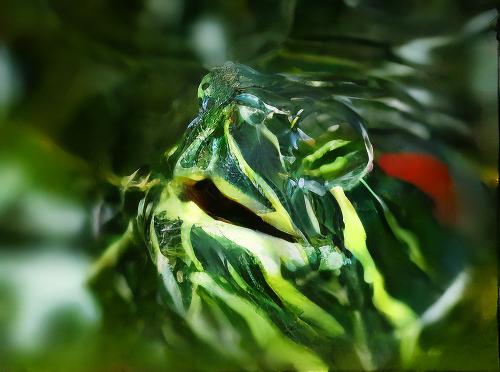}
        \caption*{Our results}
    \end{minipage}
    \begin{minipage}{0.28\textwidth}
        \centering
        \includegraphics[width=\linewidth]{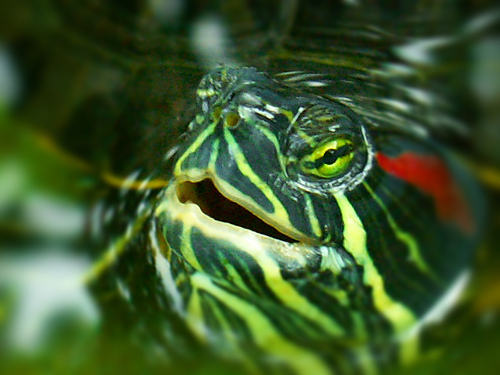}
        \caption*{Original HR}
    \end{minipage}
    
    \caption{Two representative SupResDiffGAN outputs: (top) 4× face superresolution at 128×128→512×512 pixels (bottom) 4× natural image super-resolution at 125×93→500×372 pixels.}
    \label{fig:comparison}
\end{figure*}

Image super-resolution (SR) has become an essential task in various fields, aiming to enhance the quality of low-resolution images by reconstructing finer details and improving overall clarity (Figure \ref{fig:comparison}). 

Among the different methods explored for SR, GAN-based methods \cite{ledig2017photo,wang2018esrgan,wang2021real} have demonstrated considerable success in producing high-resolution images by learning complex data distributions. GANs are particularly effective at generating realistic textures, making them a popular choice for SR tasks.

Diffusion models \cite{saharia2022image,liu20232,yue2024resshift} have recently gained attention for their ability to refine noisy images into high-quality outputs. These models work by iteratively removing noise from an image, following a learned reverse process that models the underlying data distribution. Diffusion models are particularly effective at generating fine-grained details, making them a powerful alternative to GANs for SR tasks, especially in scenarios that require highly detailed and stable textures over time, such as in applications involving time-lapse imagery or video sequences. 

Despite the advancements in both GAN-based and diffusion-based super-resolution techniques, the use of diffusion models in combination with GANs for SR remains relatively unexplored. Diffusion models offer a complementary approach to the generative capabilities of GANs. However, architectures that integrate these two powerful techniques are still rare, and there is significant potential to explore how such hybrid systems could further enhance image quality, reduce noise, and preserve critical details.

In this paper, we introduce a novel architecture that combines the strengths of GANs and diffusion models for super-resolution tasks. By incorporating a discriminator network trained adversarially within the diffusion framework, we enhance the generator’s ability to produce realistic predictions with fewer inference steps. Our approach operates in the latent space, leveraging its compressed representation to improve efficiency while maintaining high-quality outputs. By combining the fast inference time of GANs with the noise-reduction capabilities of diffusion models, we achieve superior image quality. We believe this architecture has the potential to advance research in SR and extend the capabilities of existing methodologies. In summary, our key contributions are as follows:

\begin{itemize}
    \item We introduce SupResDiffGAN, a novel SR architecture that combines the realistic texture generation of diffusion models with the high inference efficiency of GANs. Additionally, latent-space modeling further accelerates generation.
    \item To address the problem of overfitting the discriminator, we propose adaptive noise corruption, a technique that maintains a stable balance between the generator and discriminator during training.
    \item Extensive experiments on top benchmark datasets demonstrate that SupResDiffGAN outperforms existing diffusion-based super-resolution methods in image quality (measured by the LPIPS metric) and inference speed.
\end{itemize}
 
\section{Related Work}

\textbf{GANs in SR}. The introduction of GAN-based architectures in the SR task, particularly with SRGAN \cite{ledig2017photo}, represents a significant advancement over traditional CNN models trained to minimize the distance between prediction and ground truth in the pixel space \cite{dong2015image,kim2016accurate,lim2017enhanced}. A key innovation introduced by GAN-based models was the use of adversarial loss function, which, unlike conventional mathematical loss functions such as L1 and L2, prioritizes perceptual quality more than minimizing pixel-wise differences. Over time, GAN architectures in SR have evolved by improving the stability of the learning process, with innovations like Relativistic GAN loss introduced in ESRGAN \cite{wang2018esrgan}, Wasserstein GAN loss \cite{arjovsky2017wasserstein} and further refinements through fine-tuning and data augmentation techniques in ESRGAN+ \cite{rakotonirina2020esrgan+}. Subsequent iterations have expanded the capabilities of the architectures by adapting to handle real, unprocessed data, as seen in RealSR \cite{ji2020real}, Real-ESRGAN \cite{wang2021real}, BSRGAN \cite{zhang2021designing} or by adding multiple discriminators, like in MPDGAN \cite{lee2019multi}.

\textbf{Diffusion Models in SR}. Diffusion models, initially recognized for their success in image generation tasks \cite{dhariwal2021diffusion,ho2020denoising}, were later adapted for the SR, with SR3 \cite{saharia2022image} and SRDiff \cite{li2022srdiff} as precursors. Through their training stability, as well as their improved representation of detail in the image, they quickly became the new state-of-the-art techniques for the SR task (SR3+ \cite{sahak2023denoising}). Their strength lies in modeling data distribution through iterative noise removal, enabling them to capture richer details compared to earlier single-step GAN approaches \cite{ji2020real,wang2021real}.
Over time, diffusion models have evolved rapidly by improving control over the upsampling process, as seen in the Implicit Diffusion Model \cite{gao2023implicit}; modifying sampling trajectories, such as ResShift \cite{yue2024resshift} and I$^2$SB \cite{liu20232};  and accelerating inference, as demonstrated by SinSR \cite{wang2024sinsr}. More recently, efforts to push the limits of SR performance have focused on the use of the stable diffusion architecture, as evidenced by StableSR \cite{wang2024exploiting}, OmniSSR \cite{li2024omnissr} or PASD \cite{yang2023pixel}.

\textbf{GAN-Diffusion architecture}. Recent studies \cite{kuznedelev2024does} have highlighted the advantages of GAN models, indicating that they are more suitable for the SR task compared to diffusion models \cite{wang2022diffusion}. These works point to key disadvantages of diffusion models, particularly their computational complexity and long inference times \cite{xu2024ufogen}, which make them unsuitable for real-world applications \cite{wolters2023zooming}. Similar issues have been addressed in domains such as image generation, text-to-image, or image translation \cite{liu2024ladiffgan,yang2024structure}. One promising solution may be the integration of GAN and diffusion models in a Diffusion-GAN architecture \cite{wang2022diffusion,xiao2021tackling}. The approach combines the high-quality image generation of diffusion models with the faster inference speed of GANs.

Initial applications of this architecture for SR have shown promising results \cite{niu2023difauggan,xiao2024single}, but we aim to take it a step further by sampling in the latent space \cite{trinh2024latent} of the pretrained VAE model introduced in \cite{rombach2022high}. To prevent $\discriminator$ from overfitting, we employ adaptive noise augmentation to its input, following the approach introduced in \cite{wang2022diffusion}. This approach will reduce the computational complexity and improve the quality of the upsampling process.

\section{Preliminary}

Denoising Diffusion Probabilistic Models (DDPMs) \cite{ho2020denoising} are generative models that synthesize high-quality data samples by iteratively transforming random noise into meaningful data. This is achieved by reversing a forward diffusion process that progressively corrupts the data with noise. \\
Diffusion models operate as two Markov chains: the \textbf{forward process} and the \textbf{reverse process}. The forward process incrementally adds Gaussian noise to the data, degrading it to pure noise. In contrast, the reverse process learns to reconstruct the original data distribution by modeling the reverse dynamics. \\

\noindent
\textbf{The forward process} is crucial for training the diffusion model. Given a clean sample $x_0$, it creates a sequence of noisy versions of that sample, following the formula
\begin{equation}
\label{eq:forward-process}
q(x_t \mid x_{t-1}) = \mathcal{N}(x_t; \sqrt{1 - \beta_t} x_{t-1}, \beta_t \mathbb{I}),
\end{equation}
where $\beta_t$ is a noise scheduler that controls the noise level at timestep $t$. 
By repeating this formula, we get more and more noisy samples as $t$ increases, and eventually, after a sufficient number of steps $t \rightarrow T$, we get $x_T$ which should be indistinguishable from pure noise $\mathcal{N}(\mathbf{0}, \mathbb{I})$. After clever reparametrizations $\alpha_t = 1 - \beta_t; \; \bar{\alpha}_t = \prod_{s=1}^t \alpha_t$, we can marginalize all intermediate states to obtain
\begin{equation}
\label{eq:marginalized-forward}
q(x_t \mid x_0) = \mathcal{N}(x_t; \sqrt{\bar{\alpha}_t} x_0, (1 - \bar{\alpha}_t) \mathbb{I}).
\end{equation}
The main idea behind the forward process is to create training samples $(x_t, x_0)$ with any arbitrary timestep $t$, so we can train a neural network that reverses this process. \\

\noindent
\textbf{The reverse process} aims to transform pure noise $x_T \sim \mathcal{N}(0, 1)$ to clean data $x_0$. Theoretically, it is impossible because the posterior $q(x_{t-1} \mid x_t)$ is intractable. However, we can make it tractable with the additional condition on a clean sample $x_0$
\begin{align}
\label{eq:posterior}
q(x_{t-1} \mid x_t, x_0) = \mathcal{N}\left(x_{t-1}; 
\frac{\sqrt{\alpha_t}(1 - \bar{\alpha}_{t-1})}{1 - \bar{\alpha}_t} x_t + \frac{\sqrt{\bar{\alpha}_{t-1}}\beta_t}{1 - \bar{\alpha}_t} x_0,
\frac{1 - \bar{\alpha}_{t-1}}{1 - \bar{\alpha}_t} \beta_t \mathbb{I}
\right).
\end{align}
Our goal here is to propose a model $p_\theta(x_{t-1} \mid x_t)$ with its parameters $\theta$ such that minimizes Kullback-Leibler divergence between tractable posterior across all timesteps
\begin{align}
\label{eq:optimal-weights}
\theta^* = \mathop{\arg \min}\limits_{\theta} \sum_{t=1}^T D_{KL} \left(p_\theta(x_{t-1} \mid x_t) \; || \; q(x_{t-1} \mid x_t, x_0)\right),
\end{align}
which can be simplified to L2 norm between true and predicted noise $\epsilon_t$ in noisy state $x_t$. \\
Assuming a previously trained model $p_\theta(x_{t-1} \mid x_t)$, we can generate new sample, starting with $x_T \sim \mathcal{N}(0, \mathbb{I})$ and iteratively denoise one step using our model
\begin{align}
\label{eq:sampling}
p_\theta(x_{0:T}) = q(x_T) \prod_{t=1}^T p_\theta(x_{t-1} \mid x_t).
\end{align}

\begin{figure}[t]
    \includegraphics[width=\textwidth]{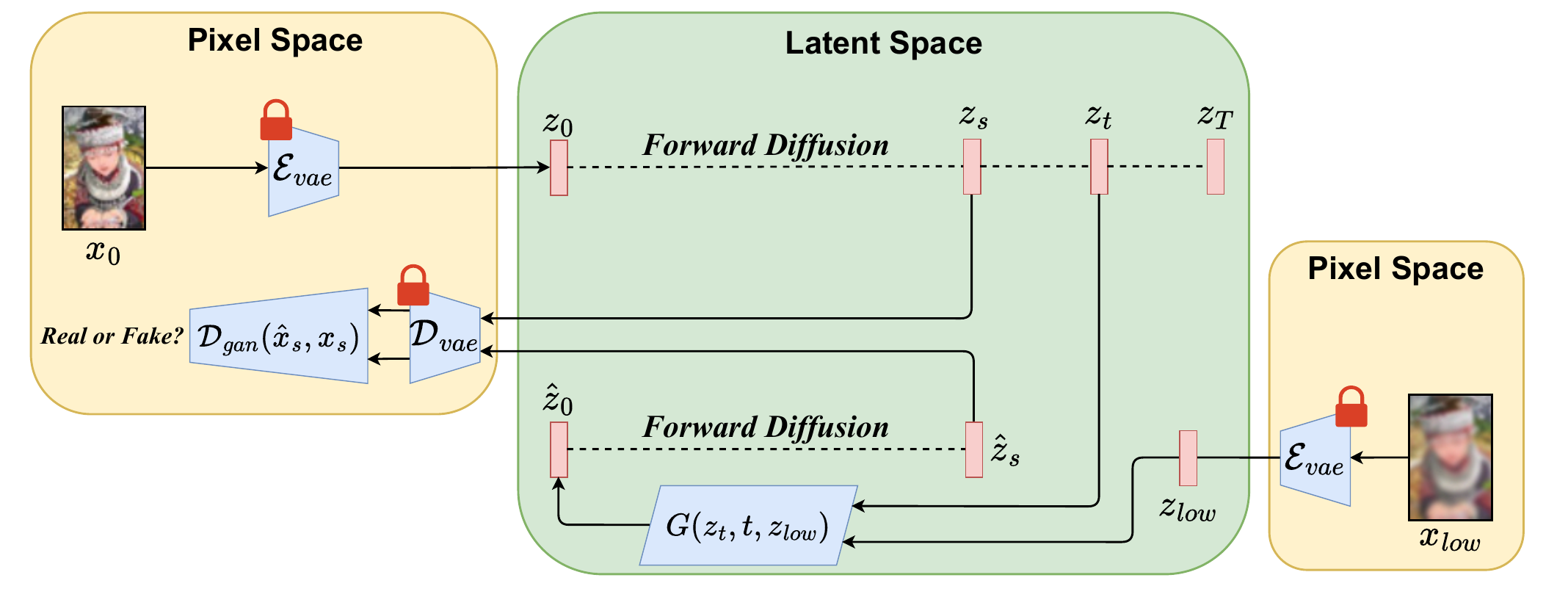}
    \caption{The training process of our proposed model. Ground truth $x_0$ and low-resolution image $x_{low}$ are embedded into latent space. The ground truth latent $z_0$ is diffused to random timestep $t$ and goes as input to the generator with low resolution latent $z_{low}$. The output of generator $\hat{z}_0$ and $z_0$ are diffused to specific timestep $s$ and decoded to pixel space where they are assessed by the discriminator which sample is real. The final loss function of the model is the mean square error between $z_0$ and $\hat{z}_0$ enriched by the adversarial loss provided by the discriminator.} 
    \label{fig:training}
\end{figure}
\section{Method}
In this section, we present SupResDiffGAN, a novel diffusion model for the SR task. Consistent with prior works, we assume that the low-resolution $x_{low}$ and high-resolution $x_0$ images have matching spatial resolutions, achieved if necessary through pre-upsampling the $x_{low}$ image using bicubic interpolation. SupResDiffGAN combines a latent space representation with a hybrid diffusion-GAN architecture, achieving a trade-off between perceptual quality and computational efficiency. \\

\noindent
\textbf{Training}
The training setup is illustrated in Figure \ref{fig:training}. The architecture of the models consists of a U-net generator $\generator$, a discriminator network $\discriminator$, and a pretrained variational autocoder (VAE) with frozen parameters. These are divided into an encoder $\encoder$ and a decoder $\decoder$. For each training pair, high-resolution $x_0$ and low-resolution $x_{low}$, we encode them first to latent representations $z_0$ and $z_{low}$ using encoder $\encoder$. This latent encoding enables efficient processing in a lower-dimensional space.

Next, We create the noised version of $z_0$, $z_t$ using the diffusion forward process from Equation \eqref{eq:marginalized-forward} at a random timestep $t$. The triplet $(z_t, t, z_{low})$ is then fed into the generator $G$, which aims to predict $\hat{z}_0$ as close to the ground-truth latent $z_0$ as possible.

To further refine the prediction, both $z_0$ and $\hat{z}_0$ undergo an additional diffusion step to the same timestep $s$, following Equation \eqref{eq:marginalized-forward}. The resulting $\hat{z}_s$ and $z_s$ are then decoded into pixel space using a pretrained decoder $\decoder$. These decoded images are passed to the discriminator $\discriminator$. The timestep $s$ is dynamically adjusted based on the discriminator's accuracy to prevent overfitting, ensuring robust training. \\

\noindent
\textbf{Generator}
The generator $\generator$, a U-Net in our case, is integrated into the diffusion process, taking as inputs: (1) the current latent state $z_t$, (2) the time step $t$, and (3) a low-resolution image in the latent space $z_{low}$ to guide generation. Its primary goal is to produce a realistic estimation of a clean latent $\hat{z}_0$ for each timestep $t$ in the diffusion reverse process
\begin{equation}
    \hat{z}_0 = \generator(z_t, t, z_{low}).
\end{equation}

\noindent
During inference, after performing diffusion sampling in the latent space, the final generated representation $\hat{z}_0$ is decoded into pixel space by the decoder $\decoder$, producing the high-resolution output image $\hat{x}_0$. \\

\noindent
\textbf{Discriminator}
The discriminator $\discriminator$ is designed to differentiate between real high-resolution images $x_0$, and super-resolution generated images $\hat{x}_0$. 

To prevent $\discriminator$ from overfitting, we introduced Gaussian noise augmentation, which utilizes the diffusion forward process in the latent space and applies it to the discriminator input
\begin{align}
    z_s \sim \mathcal{N}(\sqrt{\bar{\alpha}_s}z_0, (1 - \bar{\alpha}_s) \mathbb{I}), \\
    \hat{z}_s \sim \mathcal{N}(\sqrt{\bar{\alpha}_s}\hat{z}_0, (1 - \bar{\alpha}_s) \mathbb{I}).
\end{align}
For further stabilization of training, we employed an Exponential Moving Average (EMA) mechanism on the accuracy of the discriminator to dynamically adjust the augmentation timestep $s$ associated with the noise strength applied to the inputs
\begin{equation}
    \text{acc}_{\text{ema}}^{(i)} = \text{acc}_{\text{batch}} \cdot \lambda_{\text{ema}} + \text{acc}_{\text{ema}}^{(i-1)} \cdot (1 - \lambda_{\text{ema}}).
\end{equation}
Where $\text{acc}_{\text{ema}}^{(i)}$ and $\text{acc}_{\text{ema}}^{(i-1)}$ are EMA values on discriminator accuracy at the $i$ and $(i-1)$ training iteration. $\text{acc}_{\text{batch}}$ is the accuracy of the discriminator at the current batch, and $\lambda_{\text{ema}}$ is the EMA weight set to $0.05$. EMA monitors the discriminator's accuracy during training and adjusts $s$ with the formula
\begin{equation}
    s = \left\lfloor \max \left( 2 T \left(\text{acc}_{\text{ema}} - \frac{1}{2}\right), 0 \right) \right\rfloor, 
\end{equation}
where $T$ is the maximum diffusion timestep. This ensures that the discriminator's task is neither easy nor difficult, preventing overfitting and promoting adversarial stability. Noisy latents $z_s$ and $\hat{z}_s$ are then decoded to
$x_s = \decoder(z_s); \; \hat{x}_s = \decoder(\hat{z}_s)$.

To enhance the robustness of the adversarial training, the input of the discriminator consists of a concatenation $\oplus$ of both images $x_s$ and $\hat{x}_s$ along the channel dimension in a randomized order
\begin{equation}
    x_s \oplus_{rand} \hat{x}_s = 
    \begin{cases} 
        x_s \oplus \hat{x}_s, & \text{with probability } 0.5 \\ 
        \hat{x}_s \oplus x_s, & \text{with probability } 0.5.
    \end{cases}
\end{equation}

\noindent
The discriminator is trained to distinguish $x_s$ from $\hat{x}_s$ by predicting a binary label $y$ corresponding to the correct order of concatenation. The randomness in $\oplus_{rand}$ prevents the discriminator from relying on input order, encouraging it to focus on the details of given images instead.

The discriminator loss \eqref{eq:loss_d} is calculated using binary cross-entropy (BCE) between the predicted label and the ground truth input order. It is defined as
\begin{align}
\label{eq:loss_d}
\mathcal{L}_{D} = -\frac{1}{N} \sum_{n=1}^N \Big[ 
& y^{(n)} \log\left(\discriminator(x_s^{(n)} \oplus \hat{x}_s^{(n)})\right) \nonumber \\
& + (1 - y^{(n)}) \log\left(1 - \discriminator(\hat{x}_s^{(n)} \oplus x_s^{(n)})\right) 
\Big],
\end{align}
where $N$ is the batch size, and $(n)$ is the index of a sample. \\

\noindent
\textbf{Generator Loss}
The generator loss \( \mathcal{L}_G \) is a weighted combination of two key components: \textbf{content loss} (\( \mathcal{L}_{mse} \)) and \textbf{adversarial loss} (\( \mathcal{L}_{adv} \)). Each component serves a distinct purpose in ensuring the quality and realism of the generated images.

The \textbf{content loss} minimizes the structural difference between the generated image and the ground truth in latent space using Mean Squared Error (MSE). This term ensures that the generated image closely resembles the target image at a pixel level. For a ground truth latent $z_0$ and its generated counterpart $\hat{z}_0$, the content loss is defined as
\begin{equation}
\mathcal{L}_{mse} = || z_0 - \hat{z}_0 ||^2
\end{equation}

The \textbf{adversarial loss} encourages the generator to produce realistic textures and details by fooling the discriminator $\discriminator$. It is defined as the negative discriminator loss
\begin{equation}
\label{eq:loss_adv}
\mathcal{L}_{adv} = -\mathcal{L}_D
\end{equation}

The \textbf{total generator loss} \eqref{eq:loss_g} combines these components with their respective weights, balancing structural similarity, perceptual quality, and realism:

\begin{equation}
\label{eq:loss_g}
\mathcal{L}_G = \mathcal{L}_{mse} + \lambda_{adv} \cdot \mathcal{L}_{adv}
\end{equation}

Here, \( \lambda_{adv} = 1 \times 10^{-3} \) is a hyperparameter controlling the contributions of the adversarial losses. These weights ensure the generator prioritizes structural accuracy while enhancing perceptual appeal and realism. Balancing these terms is critical to achieving high-quality results.

The model weights are updated alternately using discriminator loss \eqref{eq:loss_d} and generator loss \eqref{eq:loss_g}. \\

\begin{figure}[t]
    \includegraphics[width=\textwidth]{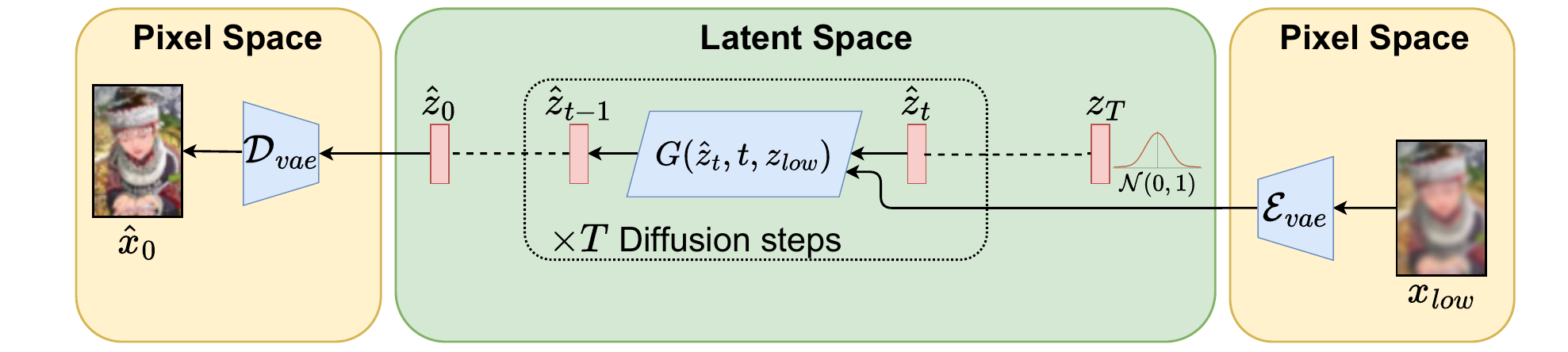}
    \caption{The sampling process of our model. We first embed the input  $x_{low}$ to the latent representation $z_{low}$. We then start the diffusion reverse process from the pure Gaussian noise and gradually remove the noise to obtain the final sample in the latent space $\hat{z}_0$ which is decoded to pixel space.}
    \label{fig:inference}
\end{figure}

\noindent
\textbf{Inference}
The inference process is illustrated in Figure \ref{fig:inference}. It begins by encoding the low-resolution input image $x_{low}$ into the latent space $z_{low}$. The diffusion reverse process, which starts from pure Gaussian noise in the latent space, incorporates the generator $\generator$ to gradually reduce the noise.

Although the generator is trained to estimate the clean latent $\hat{z}_0$, we can use Equation \eqref{eq:posterior} to obtain the estimation from subsequent timestep $\hat{z}_{t-1}$. 
Once the reverse process is complete, the final prediction $\hat{z}_0$ is decoded into pixel space using a decoder $\decoder$. 

\section{Experiments}

This section evaluates the performance and efficiency of the proposed SupResDiffGAN model architecture for single-image super-resolution (SR). The experiments aim to demonstrate that the proposed approach achieves high-quality image restoration at different scales and maintains superior perceptual and quantitative performance compared to current methods. In addition, the impact of hyperparameters and architecture elements on the model results was tested.

\textbf{Training details} We use over 400k high-resolution (HR) images from the ImageNet dataset \cite{deng2009imagenet} as the training dataset. These images are provided at various resolutions and irregular shapes ranging from 300 to 400 pixels in height and from 400 to 500 pixels in width, ensuring the adaptability and scalability of the model to diverse input sizes. To generate the corresponding low-resolution (LR) images, the HR data are downsampled by a factor of 4 using bicubic interpolation and then resized back to their original dimensions for use in the diffusion model. 
The model is trained using the Adam optimizer \cite{diederik2014adam} with a constant learning rate of 1e-4. We set the batch size to 8. To encode the input data into the latent space, we utilize a pretrained VAE from the Stable Diffusion model \cite{rombach2022high}.

\textbf{Datasets:} We conduct experiments on diverse datasets commonly used in SR research. For general testing, we use 512 images from ImageNet \cite{deng2009imagenet}, aligned with training specifications. CelebA-HQ \cite{karras2017progressive} is used for evaluating fine-grained facial details. Div2K \cite{agustsson2017ntire}, and Urban100 \cite{huang2015single} assess adaptability across resolutions and detail recovery. Real-world degradation is tested with RealSR \cite{cai2019toward}, containing images from Canon 5D3 and Nikon D810. Finally, we included Set14 \cite{zeyde2012single}, a widely-used but small benchmark. This selection ensures a comprehensive evaluation across synthetic, high-detail, and real-world scenarios.

\textbf{Methods:} Several state-of-the-art Super-Resolution (SR) methods are selected as baselines for comparison. To evaluate the performance of the proposed model against GAN-based approaches, we include SRGAN \cite{ledig2017photo}, ESRGAN \cite{wang2018esrgan}, and Real-ESRGAN \cite{wang2021real}. For diffusion-based methods, comparisons are made with SR3 \cite{saharia2022image}, ResShift \cite{yue2024resshift}, and I$^2$SB \cite{liu20232}.

\textbf{Metrics:} The model was evaluated using quantitative and qualitative indicators. PSNR and SSIM \cite{wang2004image} were used to assess fidelity, while LPIPS \cite{zhang2018unreasonable} was used to assess perceptual quality and realism. Additionally, the time needed to generate a batch of images was measured to evaluate the model's suitability for real-world applications. All methods were evaluated under the same conditions using an NVIDIA A100 GPU and the same batch size to ensure fair comparisons.

\begin{center}
\begin{table*}[t]
\caption{Comparison of methods on CelebA-HQ dataset across 5 metrics: LPIPS, SSIM, PSNR, MSE and Time per batch [s]. The best and second best results are highlighted in \textbf{bold} and \underline{underline}. Methods are categorized into \textit{Diffusion-based} and \textit{GAN-based} to reflect their distinct architectural frameworks.}
\begin{center}
\begin{tabular}{|c|c|c|c|c|c|}
\hline
\textit{\textbf{Metric}}        & \textit{\textbf{Lpips}} \(\downarrow\)  & \textit{\textbf{SSIM}} \(\uparrow\)   & \textit{\textbf{PSNR}} \(\uparrow\)    & \textit{\textbf{MSE}} \(\downarrow\)      & \textit{\textbf{Time per batch {[}s{]}}} \(\downarrow\) \\ \hline
\multicolumn{6}{|c|}{\textit{\textbf{GAN-based methods}}}                                                                                                                                                                                                                                \\ \hline
\multicolumn{1}{|c|}{\textit{\textbf{SRGAN}}}         & \multicolumn{1}{c|}{0.2441}                  & \multicolumn{1}{c|}{\textbf{0.8186}}        & \multicolumn{1}{c|}{\textbf{27.8723}}       & \multicolumn{1}{c|}{\textbf{0.0017}}     & \textbf{0.0109}                          \\ \hline
\multicolumn{1}{|c|}{\textit{\textbf{ESRGAN}}}        & \multicolumn{1}{c|}{{\underline{0.1903}}}            & \multicolumn{1}{c|}{0.6844}                 & \multicolumn{1}{c|}{23.4248}                & \multicolumn{1}{c|}{0.0047}              & 0.0870                                    \\ \hline
\multicolumn{1}{|c|}{\textit{\textbf{Real-ESRGAN}}}   & \multicolumn{1}{c|}{\textbf{0.1690}}          & \multicolumn{1}{c|}{{\underline{0.7426}}}           & \multicolumn{1}{c|}{{\underline{26.2697}}}          & \multicolumn{1}{c|}{{\underline{0.0025}}}        & {\underline{0.0816}}                             \\ \hline
\multicolumn{6}{|c|}{\textit{\textbf{Diffusion-based methods}}}                                                                                                                                                                                                                          \\ \hline
\multicolumn{1}{|c|}{\textit{\textbf{SR3}}}           & \multicolumn{1}{c|}{0.2229}                  & \multicolumn{1}{c|}{\textbf{0.8149}}        & \multicolumn{1}{c|}{\textbf{28.0799}}       & \multicolumn{1}{c|}{\textbf{0.0016}}       & 0.3072                                   \\ \hline
\multicolumn{1}{|c|}{\textit{\textbf{I2SB}}}          & \multicolumn{1}{c|}{\underline{0.2221}}         & \multicolumn{1}{c|}{{\underline{0.7990}}}            & \multicolumn{1}{c|}{{\underline{27.2533}}}          & \multicolumn{1}{c|}{{\underline{0.0020}}}        & \textbf{0.1184}                          \\ \hline
\multicolumn{1}{|c|}{\textit{\textbf{ResShift}}}      & \multicolumn{1}{c|}{0.3275}                  & \multicolumn{1}{c|}{0.7254}                 & \multicolumn{1}{c|}{23.5016}                & \multicolumn{1}{c|}{0.0047}              & 0.4394                                   \\ \hline
\multicolumn{1}{|c|}{\textit{\textbf{SupResDiffGAN}}} & \multicolumn{1}{c|}{\textbf{0.1875}}         & \multicolumn{1}{c|}{0.7485}                 & \multicolumn{1}{c|}{26.1134}                & \multicolumn{1}{c|}{0.0026}              & {\underline{0.1832}}                             \\ \hline
\end{tabular}
\end{center}
\label{tab:all_metrics}
\end{table*}
\end{center}
\subsection{Comparison with State-of-the-Art Methods}

\textbf{Experimental setup:} All seven tested methods, including SupResDiff, were trained on ImageNet with 330,000 steps and a batch size of 8, ensuring fair comparison. The final checkpoint is selected based on the best LPIPS score on the validation set. Diffusion-based methods share the same U-Net architecture (50M), while GANs retain their original implementations. For evaluation, all diffusion models use a step size of 10, except for CelebA-HQ, where a step size of 3 is used due to easier modality. Identical datasets and batch settings ensure consistency, allowing an objective assessment of efficiency and quality across models.

\begin{center}
\begin{table*}[h]
\caption{Comparison of LPIPS metric across all evaluation datasets. The best and second best results are highlighted in \textbf{bold} and \underline{underline}. Methods are categorized into \textit{Diffusion-based} and \textit{GAN-based} to reflect their distinct architectural frameworks.}
\begin{center}
\begin{tabular}{|c|ccccccc|}
\hline
\multicolumn{1}{|c|}{\textbf{Model / Dataset}}        & \multicolumn{1}{c|}{\textit{\textbf{Imagenet}}} & \multicolumn{1}{c|}{\textit{\textbf{Celeb}}} & \multicolumn{1}{c|}{\textit{\textbf{Div2k}}} & \multicolumn{1}{c|}{\textit{\textbf{RealSR-nikon}}} & \multicolumn{1}{c|}{\textit{\textbf{RealSR-canon}}} & \multicolumn{1}{c|}{\textit{\textbf{Set14}}} & \textit{\textbf{Urban100}} \\ \hline
\multicolumn{1}{|c|}{\textit{\textbf{Metric}}}        & \multicolumn{7}{c|}{\textit{\textbf{LPIPS} \(\downarrow\)}}                                                                                                                                                                                                                                                                                          \\ \hline
\multicolumn{8}{|c|}{\textit{\textbf{GAN-based methods}}}                                                                                                                                                                                                                                                                                                                                     \\ \hline
\multicolumn{1}{|c|}{\textit{\textbf{SRGAN}}}         & \multicolumn{1}{c|}{0.3452}                     & \multicolumn{1}{c|}{0.2441}                  & \multicolumn{1}{c|}{0.3327}                  & \multicolumn{1}{c|}{0.3464}                         & \multicolumn{1}{c|}{{\underline{0.3050}}}                    & \multicolumn{1}{c|}{0.2901}                  & 0.3156                     \\ \hline
\multicolumn{1}{|c|}{\textit{\textbf{ESRGAN}}}        & \multicolumn{1}{c|}{{\underline{0.2320}}}                & \multicolumn{1}{c|}{{\underline{0.1903}}}            & \multicolumn{1}{c|}{{\underline{0.2649}}}            & \multicolumn{1}{c|}{{\underline{0.3380}}}                    & \multicolumn{1}{c|}{0.3053}                         & \multicolumn{1}{c|}{{\underline{0.2375}}}            & {\underline{0.2408}}               \\ \hline
\multicolumn{1}{|c|}{\textit{\textbf{Real-ESRGAN}}}   & \multicolumn{1}{c|}{\textbf{0.2123}}            & \multicolumn{1}{c|}{\textbf{0.1690}}          & \multicolumn{1}{c|}{\textbf{0.2562}}         & \multicolumn{1}{c|}{\textbf{0.3309}}                & \multicolumn{1}{c|}{\textbf{0.3020}}                 & \multicolumn{1}{c|}{\textbf{0.2301}}         & \textbf{0.2285}            \\ \hline
\multicolumn{8}{|c|}{\textit{\textbf{Diffusion-based methods}}}                                                                                                                                                                                                                                                                                                                               \\ \hline
\multicolumn{1}{|c|}{\textit{\textbf{SR3}}}           & \multicolumn{1}{c|}{{\underline{0.3519}}}               & \multicolumn{1}{c|}{0.2229}                  & \multicolumn{1}{c|}{0.3396}                  & \multicolumn{1}{c|}{{\underline{0.4018}}}                   & \multicolumn{1}{c|}{0.4008}                         & \multicolumn{1}{c|}{{\underline{0.3015}}}            & \textbf{0.2428}                     \\ \hline
\multicolumn{1}{|c|}{\textit{\textbf{I$^2$SB}}}          & \multicolumn{1}{c|}{0.3755}                     & \multicolumn{1}{c|}{{\underline{0.2221}}}            & \multicolumn{1}{c|}{{\underline{0.3309}}}            & \multicolumn{1}{c|}{0.4069}                         & \multicolumn{1}{c|}{{\underline{0.3867}}}                   & \multicolumn{1}{c|}{0.3169}                  & {0.2635}               \\ \hline
\multicolumn{1}{|c|}{\textit{\textbf{ResShift}}}      & \multicolumn{1}{c|}{0.5360}                      & \multicolumn{1}{c|}{0.3275}                  & \multicolumn{1}{c|}{0.4724}                  & \multicolumn{1}{c|}{0.4959}                         & \multicolumn{1}{c|}{0.4671}                         & \multicolumn{1}{c|}{0.4832}                  & 0.4822                     \\ \hline
\multicolumn{1}{|c|}{\textit{\textbf{SupResDiffGAN}}} & \multicolumn{1}{c|}{\textbf{0.3079}}            & \multicolumn{1}{c|}{\textbf{0.1875}}         & \multicolumn{1}{c|}{\textbf{0.2876}}         & \multicolumn{1}{c|}{\textbf{0.3970}}                 & \multicolumn{1}{c|}{\textbf{0.3853}}                & \multicolumn{1}{c|}{\textbf{0.2789}}         & \underline{0.2570}             \\ \hline
\end{tabular}
\end{center}
\label{tab:lpips}
\end{table*}
\end{center}


\textbf{Quality Comparison:} As shown in Tables \ref{tab:all_metrics} and \ref{tab:lpips}, GAN-based methods, particularly ESRGAN and Real-ESRGAN, achieve the best LPIPS performance across all datasets. While the proposed method performs slightly worse than these, it significantly improves upon other diffusion-based models, achieving the best results—including SR3, a purely diffusion-based approach to super-resolution. 
Although our method does not perform as well in terms of PSNR and SSIM, these metrics tend to favor overly smooth results.


The relatively lower performance of diffusion models in this study compared to their original benchmarks is due to the fact, that both GANs and diffusion models share similar model sizes. Since diffusion models typically require more parameters and converge more slowly, their performance is constrained under these conditions.
Despite being outperformed by GANs in this setting, the proposed architecture demonstrates strong potential. Enhancing diffusion models while incorporating adversarial elements serves as a promising bridge between GANs and diffusion-based approaches, with the potential to surpass state-of-the-art methods in the future.

\begin{center}
\begin{table*}[h]
\caption{Comparison of time of the batch inference in seconds. The best and second best results are highlighted in \textbf{bold} and \underline{underline}. Methods are categorized into \textit{Diffusion-based} and \textit{GAN-based} to reflect their distinct architectural frameworks.}
\begin{center}
\begin{tabular}{|c|ccccccc|}
\hline
\multicolumn{1}{|c|}{\textbf{Model / Dataset}}        & \multicolumn{1}{c|}{\textit{\textbf{Imagenet}}} & \multicolumn{1}{c|}{\textit{\textbf{Celeb}}} & \multicolumn{1}{c|}{\textit{\textbf{Div2k}}} & \multicolumn{1}{c|}{\textit{\textbf{RealSR-nikon}}} & \multicolumn{1}{c|}{\textit{\textbf{RealSR-canon}}} & \multicolumn{1}{c|}{\textit{\textbf{Set14}}} & \textit{\textbf{Urban100}} \\ \hline
\multicolumn{1}{|c|}{\textit{\textbf{Metric}}}        & \multicolumn{7}{c|}{\textit{\textbf{Time per batch {[}s{]}}}}                                                                                                                                                                                                                                                                         \\ \hline
\multicolumn{8}{|c|}{\textit{\textbf{GAN-based methods}}}                                                                                                                                                                                                                                                                                                                                     \\ \hline
\multicolumn{1}{|c|}{\textit{\textbf{SRGAN}}}         & \multicolumn{1}{c|}{\textbf{0.0671}}            & \multicolumn{1}{c|}{\textbf{0.0109}}         & \multicolumn{1}{c|}{\textbf{0.0193}}         & \multicolumn{1}{c|}{\textbf{0.0367}}                & \multicolumn{1}{c|}{\textbf{0.0113}}                & \multicolumn{1}{c|}{\textbf{0.0888}}         & \textbf{0.0070}             \\ \hline
\multicolumn{1}{|c|}{\textit{\textbf{ESRGAN}}}        & \multicolumn{1}{c|}{0.2188}                     & \multicolumn{1}{c|}{0.0870}                   & \multicolumn{1}{c|}{0.2316}                  & \multicolumn{1}{c|}{0.2711}                         & \multicolumn{1}{c|}{0.1504}                         & \multicolumn{1}{c|}{{\underline{0.2049}}}            & {\underline{0.0821}}               \\ \hline
\multicolumn{1}{|c|}{\textit{\textbf{Real-ESRGAN}}}   & \multicolumn{1}{c|}{{\underline{0.1392}}}               & \multicolumn{1}{c|}{{\underline{0.0816}}}            & \multicolumn{1}{c|}{{\underline{0.1899}}}            & \multicolumn{1}{c|}{{\underline{0.2468}}}                   & \multicolumn{1}{c|}{{\underline{0.1427}}}                   & \multicolumn{1}{c|}{0.2361}                  & 0.1013                     \\ \hline
\multicolumn{8}{|c|}{\textit{\textbf{Diffusion-based methods}}}                                                                                                                                                                                                                                                                                                                               \\ \hline
\multicolumn{1}{|c|}{\textit{\textbf{SR3}}}           & \multicolumn{1}{c|}{1.9953}                     & \multicolumn{1}{c|}{0.3072}                  & \multicolumn{1}{c|}{7.6377}                  & \multicolumn{1}{c|}{8.4242}                         & \multicolumn{1}{c|}{3.6420}                          & \multicolumn{1}{c|}{0.8627}                  & 1.5028                     \\ \hline
\multicolumn{1}{|c|}{\textit{\textbf{I$^2$SB}}}          & \multicolumn{1}{c|}{{\underline{1.6776}}}               & \multicolumn{1}{c|}{\textbf{0.1184}}         & \multicolumn{1}{c|}{{\underline{6.7292}}}            & \multicolumn{1}{c|}{{\underline{7.0910}}}                    & \multicolumn{1}{c|}{{\underline{3.1629}}}                   & \multicolumn{1}{c|}{1.8049}                  & {\underline{1.2395}}               \\ \hline
\multicolumn{1}{|c|}{\textit{\textbf{ResShift}}}      & \multicolumn{1}{c|}{2.2466}                     & \multicolumn{1}{c|}{0.4394}                  & \multicolumn{1}{c|}{8.6647}                  & \multicolumn{1}{c|}{8.9677}                         & \multicolumn{1}{c|}{4.1880}                          & \multicolumn{1}{c|}{{\underline{0.5983}}}            & 1.6762                     \\ \hline
\multicolumn{1}{|c|}{\textit{\textbf{SupResDiffGAN}}} & \multicolumn{1}{c|}{\textbf{0.2954}}            & \multicolumn{1}{c|}{{\underline{0.1832}}}            & \multicolumn{1}{c|}{\textbf{0.9333}}         & \multicolumn{1}{c|}{\textbf{1.0021}}                & \multicolumn{1}{c|}{\textbf{0.6114}}                & \multicolumn{1}{c|}{\textbf{0.3542}}         & \textbf{0.3206}            \\ \hline
\end{tabular}
\end{center}
\label{tab:time}
\end{table*}
\end{center}

\textbf{Efficiency Comparison:} Table \ref{tab:time} compares inference times per batch across different models and datasets, highlighting the computational efficiency of SupResDiffGAN. While GAN-based methods like SRGAN and ESRGAN are known for their fast inference speeds, diffusion models such as SR3 and I$^2$SB typically suffer from significantly longer times due to their iterative denoising process. In contrast, SupResDiffGAN achieves inference speeds comparable to ESRGAN, making it highly suitable for real-world applications. This efficiency results from two key factors: (1) Latent Space Representation: By operating in a reduced-dimensional latent space, SupResDiffGAN accelerates the diffusion process. (2) Reduced Diffusion Steps: Leveraging the latent space, SupResDiffGAN generates high-quality images in as few as 10 steps, compared to the thousands required by traditional diffusion models.

\begin{figure}[t]
    \centering
    \begin{adjustbox}{minipage=1.0\textwidth, valign=t, scale=1}
        \begin{minipage}[t]{0.47\textwidth}  
            \centering
            \includegraphics[width=\textwidth]{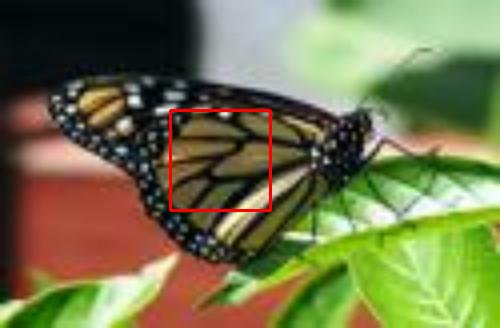}
            \includegraphics[width=\textwidth]{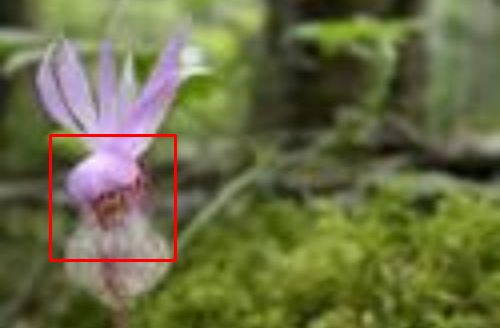}
        \end{minipage}
        \hfill
        \begin{minipage}[t]{0.512\textwidth}  
            \centering
            \vspace*{-11.6em}
            \begin{tabular}{cccc}
                \includegraphics[width=0.22\textwidth]{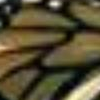} &
                \includegraphics[width=0.22\textwidth]{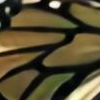} &
                \includegraphics[width=0.22\textwidth]{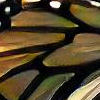} &
                \includegraphics[width=0.22\textwidth]{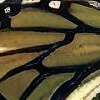} \\
                \scriptsize{Bicubic} & \scriptsize{SRGAN} & \scriptsize{ESRGAN} & \scriptsize{Real ESRGAN} \\
                
                \includegraphics[width=0.22\textwidth]{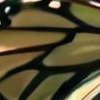} &
                \includegraphics[width=0.22\textwidth]{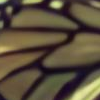} &
                \includegraphics[width=0.22\textwidth]{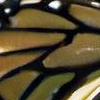} &
                \includegraphics[width=0.22\textwidth]{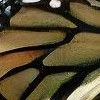} \\
                \vspace*{0.2em}
                \scriptsize{SR3} & \scriptsize{ResShift} & \scriptsize{I$^2$SB} & \scriptsize{Ours} \\
                \includegraphics[width=0.22\textwidth]{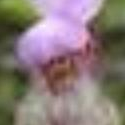} &
                \includegraphics[width=0.22\textwidth]{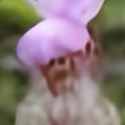} &
                \includegraphics[width=0.22\textwidth]{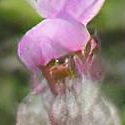} &
                \includegraphics[width=0.22\textwidth]{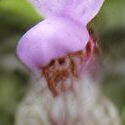} \\
                \scriptsize{Bicubic} & \scriptsize{SRGAN} & \scriptsize{ESRGAN} & \scriptsize{Real ESRGAN} \\
                \includegraphics[width=0.22\textwidth]{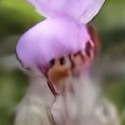} &
                \includegraphics[width=0.22\textwidth]{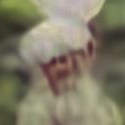} &
                \includegraphics[width=0.22\textwidth]{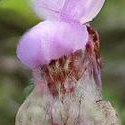} &
                \includegraphics[width=0.22\textwidth]{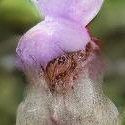} \\
                \scriptsize{SR3} & \scriptsize{ResShift} & \scriptsize{I$^2$SB} & \scriptsize{Ours} 
            \end{tabular}
        \end{minipage}
    \end{adjustbox}
    \caption{Qualitative comparison of visual performance on two example images from ImageNet. Low-quality inputs are on the left, while results from bicubic upscale and seven SR models: SRGAN, ESRGAN, Real-ESRGAN, SR3, ResShift, I$^2$SB, and Ours are on the right.}
    \label{fig:comparison_details}
\end{figure}

\textbf{Visual Performance:} Figure \ref{fig:comparison_details} compares the performance of SupResDiffGAN with GAN-based models (ESRGAN, RealESRGAN, SRGAN), diffusion methods (SR3, I$^2$SB, ResShift) and Bicubic Resize. While GAN-based models can preserve a great degree of reference information, they have a lower ability to reconstruct more detailed parts that often appear blurry, as can be observed in the bottom part of the flower image. Similarly, SR3 and ResShift fail to reproduce complex details, leading to overly smooth outputs. In contrast, models like I$^2$SB and our approach generate highly detailed super-resolution images. SupResDiffGAN leverages the strengths of both GANs and diffusion models, achieving sharp, accurate results with minimal artifacts by balancing information retention and fine-detail generation.

\begin{center}
\begin{table*}[h]
\caption{Ablation study on the influence of the GAN component in the SupResDiffGAN architecture. Evaluation metrics on the CelebA-HQ dataset comparing (1) a model without a discriminator or adversarial loss, 2) a model with a discriminator but without Gaussian noise augmentation, and (3) the full proposed architecture.}
\begin{center}
\begin{tabular}{|c|c|c|c|c|c|}
\hline
\textit{\textbf{Metric}}                 & \textit{\textbf{Lpips} \(\downarrow\)} & \textit{\textbf{SSIM} \(\uparrow\)} & \textit{\textbf{PSNR} \(\uparrow\)} & \textit{\textbf{MSE} \(\downarrow\)} \\ \hline
\textit{\textbf{(1) SupResDiffGAN - no adv}} & 0.1620                   & 0.7377                 & 25.9726                & 0.0027                          \\ \hline
\textit{\textbf{(2) SupResDiffGAN - without noise}} & 0.1591                  & 0.7363                 & 25.7346                & 0.0028                                     \\ \hline
\textit{\textbf{(3) SupResDiffGAN - our}}    & \textbf{0.1537}         & \textbf{0.7527}        & \textbf{26.2138}       & \textbf{0.0026}                                  \\ \hline
\end{tabular}
\end{center}
\label{tab:ablation}
\end{table*}
\end{center}

\subsection{Ablation Studies} 

In this section, we conduct a comprehensive analysis of the proposed GAN-Diffusion model to evaluate the impact of key architectural components. Specifically, we examine: (1) the role of \textit{adversarial loss} to determine whether it enhances model performance, (2) the effect of the number of denoising steps and the influence of different sampling strategies.




\textbf{Influence of GAN Component}
Adversarial loss plays a crucial role in our model by integrating the GAN framework within the diffusion process, as detailed in Section 4. To evaluate its impact on output quality, we conduct an experiment comparing three configurations: (1) a model without a discriminator or adversarial loss, (2) a model with a discriminator but without Gaussian noise augmentation, and (3) the full proposed architecture incorporating a discriminator, adaptive noise, and EMA. All models are trained under identical conditions on CelebA-HQ for the same number of generator learning steps, with performance assessed using 10 inference steps on the CelebA-HQ test set.

As shown in Table \ref{tab:ablation}, the full model (3) achieves the best overall performance. Interestingly, while the model without GAN (1) surpasses the simple GAN setup (2) in SSIM, PSNR, and MSE, configuration (2) achieves better LPIPS scores, underscoring the role of adversarial loss in enhancing perceptual quality. The full model (3), which combines adversarial learning and EMA, offers the best balance—improving realism and perceptual fidelity while maintaining strong structural similarity. These results highlight the effectiveness of integrating adversarial loss within the diffusion framework, demonstrating its ability to refine the model’s output quality without compromising efficiency.

\begin{figure}[t]
    \centering 
    \includegraphics[width=0.8\textwidth]{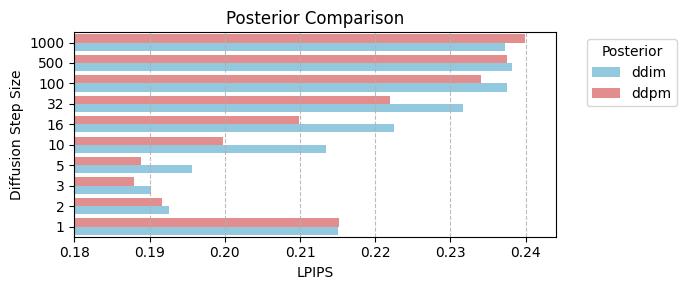}
    \caption{Impact of diffusion step size and sampling method on SupResDiffGAN performance, evaluated using LPIPS. Results are based on the CelebA-HQ dataset. The model maintained quality with fewer steps, significantly reducing inference time. DDPM outperformed DDIM at low step counts, but their results converged with more steps.}
    \label{fig:ablation_prior_steps}
\end{figure}

\textbf{Number of steps and stochasticity of the inference}
After training the model, we evaluated its performance in a different number of sampling steps using the DDPM \cite{ho2020denoising} and DDIM \cite{song2020denoising} methods. We used the LPIPS metric due to its strong correlation with human perception. As shown in Figure \ref{fig:ablation_prior_steps}, the model performed well even with very few steps (e.g. $3$ steps), reducing inference time by up to $\times 200$ compared to the standard $1000$-step process while preserving perceptual quality and fidelity. By using adversarial training, the model learned to generate more realistic outputs, reducing the need for many sampling steps. DDPM performed slightly better than DDIM at lower step counts (1–10), but the difference diminished as steps increased.

\section{Limitations and Conclusions}
Thanks to the adversarial term enhancing the diffusion process, our model achieves significant improvements in both efficiency and quality (measured by LPIPS) over traditional diffusion-based methods. However, it still has some limitations. First, due to the strong influence of the diffusion loss term, the model requires a paired dataset for training. Second, while the adversarial term enables high performance with fewer diffusion steps—leading to faster inference—the overall training time may be longer, as the discriminator must be trained alongside the generator.

The proposed SupResDiffGAN bridges GAN-based and diffusion-based super-resolution, balancing efficiency and quality. It employs adaptive noise corruption to prevent discriminator overfitting and leverages latent space representations along with fewer diffusion steps to achieve GAN-like inference speeds. Though it doesn't surpass top GAN-based models in some metrics, its strong performance highlights its potential. This work paves the way for future hybrid architectures, emphasizing scalability, interpretability, and real-world robustness, bringing SR models closer to practical adoption.

\subsection*{Acknowledgment}

\footnotesize{
We gratefully acknowledge Polish high-performance computing infrastructure PLGrid (HPC Center: ACK Cyfronet AGH) for providing computer facilities and support within computational grant no. PLG/2024/017703.

\noindent
The work conducted by Wojciech Kozłowski and Maciej Zięba was supported by the National Centre of Science (Poland) grant no. 2021/43/B/ST6/02853. 
}

\bibliography{bibliography}

\begin{thebibliography}{10}
\providecommand{\url}[1]{\texttt{#1}}
\providecommand{\urlprefix}{URL }
\providecommand{\doi}[1]{https://doi.org/#1}

\bibitem{agustsson2017ntire}
Agustsson, E., Timofte, R.: Ntire 2017 challenge on single image super-resolution: Dataset and study. In: Proceedings of the IEEE conference on computer vision and pattern recognition workshops. pp. 126--135 (2017)

\bibitem{arjovsky2017wasserstein}
Arjovsky, M., Chintala, S., Bottou, L.: Wasserstein generative adversarial networks. In: International conference on machine learning. pp. 214--223. PMLR (2017)

\bibitem{cai2019toward}
Cai, J., Zeng, H., Yong, H., Cao, Z., Zhang, L.: Toward real-world single image super-resolution: A new benchmark and a new model. In: Proceedings of the IEEE/CVF international conference on computer vision. pp. 3086--3095 (2019)

\bibitem{deng2009imagenet}
Deng, J., Dong, W., Socher, R., Li, L.J., Li, K., Fei-Fei, L.: Imagenet: A large-scale hierarchical image database. In: 2009 IEEE conference on computer vision and pattern recognition. pp. 248--255. Ieee (2009)

\bibitem{dhariwal2021diffusion}
Dhariwal, P., Nichol, A.: Diffusion models beat gans on image synthesis. Advances in neural information processing systems  \textbf{34},  8780--8794 (2021)

\bibitem{diederik2014adam}
Diederik, P.K.: Adam: A method for stochastic optimization. (No Title)  (2014)

\bibitem{dong2015image}
Dong, C., Loy, C.C., He, K., Tang, X.: Image super-resolution using deep convolutional networks. IEEE transactions on pattern analysis and machine intelligence  \textbf{38}(2),  295--307 (2015)

\bibitem{gao2023implicit}
Gao, S., Liu, X., Zeng, B., Xu, S., Li, Y., Luo, X., Liu, J., Zhen, X., Zhang, B.: Implicit diffusion models for continuous super-resolution. In: Proceedings of the IEEE/CVF conference on computer vision and pattern recognition. pp. 10021--10030 (2023)

\bibitem{ho2020denoising}
Ho, J., Jain, A., Abbeel, P.: Denoising diffusion probabilistic models. Advances in neural information processing systems  \textbf{33},  6840--6851 (2020)

\bibitem{huang2015single}
Huang, J.B., Singh, A., Ahuja, N.: Single image super-resolution from transformed self-exemplars. In: Proceedings of the IEEE conference on computer vision and pattern recognition. pp. 5197--5206 (2015)

\bibitem{ji2020real}
Ji, X., Cao, Y., Tai, Y., Wang, C., Li, J., Huang, F.: Real-world super-resolution via kernel estimation and noise injection. In: proceedings of the IEEE/CVF conference on computer vision and pattern recognition workshops. pp. 466--467 (2020)

\bibitem{karras2017progressive}
Karras, T.: Progressive growing of gans for improved quality, stability, and variation. arXiv preprint arXiv:1710.10196  (2017)

\bibitem{kim2016accurate}
Kim, J., Lee, J.K., Lee, K.M.: Accurate image super-resolution using very deep convolutional networks. In: Proceedings of the IEEE conference on computer vision and pattern recognition. pp. 1646--1654 (2016)

\bibitem{kuznedelev2024does}
Kuznedelev, D., Startsev, V., Shlenskii, D., Kastryulin, S.: Does diffusion beat gan in image super resolution? arXiv preprint arXiv:2405.17261  (2024)

\bibitem{ledig2017photo}
Ledig, C., Theis, L., Husz{\'a}r, F., Caballero, J., Cunningham, A., Acosta, A., Aitken, A., Tejani, A., Totz, J., Wang, Z., et~al.: Photo-realistic single image super-resolution using a generative adversarial network. In: Proceedings of the IEEE conference on computer vision and pattern recognition. pp. 4681--4690 (2017)

\bibitem{lee2019multi}
Lee, O.Y., Shin, Y.H., Kim, J.O.: Multi-perspective discriminators-based generative adversarial network for image super resolution. IEEE Access  \textbf{7},  136496--136510 (2019)

\bibitem{li2022srdiff}
Li, H., Yang, Y., Chang, M., Chen, S., Feng, H., Xu, Z., Li, Q., Chen, Y.: Srdiff: Single image super-resolution with diffusion probabilistic models. Neurocomputing  \textbf{479},  47--59 (2022)

\bibitem{li2024omnissr}
Li, R., Sheng, X., Li, W., Zhang, J.: Omnissr: Zero-shot omnidirectional image super-resolution using stable diffusion model. arXiv preprint arXiv:2404.10312  (2024)

\bibitem{lim2017enhanced}
Lim, B., Son, S., Kim, H., Nah, S., Mu~Lee, K.: Enhanced deep residual networks for single image super-resolution. In: Proceedings of the IEEE conference on computer vision and pattern recognition workshops. pp. 136--144 (2017)

\bibitem{liu20232}
Liu, G.H., Vahdat, A., Huang, D.A., Theodorou, E.A., Nie, W., Anandkumar, A.: {I$^2$SB: Image-to-Image Schr\"odinger Bridge} (2023), \url{https://arxiv.org/abs/2302.05872}

\bibitem{liu2024ladiffgan}
Liu, X., Zeng, B., Gao, S., Li, S., Feng, Y., Li, H., Liu, B., Liu, J., Zhang, B.: Ladiffgan: Training gans with diffusion supervision in latent spaces. In: Proceedings of the IEEE/CVF Conference on Computer Vision and Pattern Recognition. pp. 1115--1125 (2024)

\bibitem{niu2023difauggan}
Niu, A., Zhang, K., Tee, J.T.J., Pham, T.X., Sun, J., Yoo, C.D., Kweon, I.S., Zhang, Y.: Difauggan: A practical diffusion-style data augmentation for gan-based single image super-resolution. arXiv preprint arXiv:2311.18508  (2023)

\bibitem{rakotonirina2020esrgan+}
Rakotonirina, N.C., Rasoanaivo, A.: Esrgan+: Further improving enhanced super-resolution generative adversarial network. In: ICASSP 2020-2020 IEEE International Conference on Acoustics, Speech and Signal Processing (ICASSP). pp. 3637--3641. IEEE (2020)

\bibitem{rombach2022high}
Rombach, R., Blattmann, A., Lorenz, D., Esser, P., Ommer, B.: High-resolution image synthesis with latent diffusion models. In: Proceedings of the IEEE/CVF conference on computer vision and pattern recognition. pp. 10684--10695 (2022)

\bibitem{sahak2023denoising}
Sahak, H., Watson, D., Saharia, C., Fleet, D.: Denoising diffusion probabilistic models for robust image super-resolution in the wild. arXiv preprint arXiv:2302.07864  (2023)

\bibitem{saharia2022image}
Saharia, C., Ho, J., Chan, W., Salimans, T., Fleet, D.J., Norouzi, M.: Image super-resolution via iterative refinement. IEEE transactions on pattern analysis and machine intelligence  \textbf{45}(4),  4713--4726 (2022)

\bibitem{song2020denoising}
Song, J., Meng, C., Ermon, S.: Denoising diffusion implicit models. arXiv preprint arXiv:2010.02502  (2020)

\bibitem{trinh2024latent}
Trinh, L.T., Hamagami, T.: Latent denoising diffusion gan: Faster sampling, higher image quality. IEEE Access  (2024)

\bibitem{wang2024exploiting}
Wang, J., Yue, Z., Zhou, S., Chan, K.C., Loy, C.C.: Exploiting diffusion prior for real-world image super-resolution. International Journal of Computer Vision pp. 1--21 (2024)

\bibitem{wang2021real}
Wang, X., Xie, L., Dong, C., Shan, Y.: Real-esrgan: Training real-world blind super-resolution with pure synthetic data. In: Proceedings of the IEEE/CVF international conference on computer vision. pp. 1905--1914 (2021)

\bibitem{wang2018esrgan}
Wang, X., Yu, K., Wu, S., Gu, J., Liu, Y., Dong, C., Qiao, Y., Change~Loy, C.: Esrgan: Enhanced super-resolution generative adversarial networks. In: Proceedings of the European conference on computer vision (ECCV) workshops. pp.~0--0 (2018)

\bibitem{wang2024sinsr}
Wang, Y., Yang, W., Chen, X., Wang, Y., Guo, L., Chau, L.P., Liu, Z., Qiao, Y., Kot, A.C., Wen, B.: Sinsr: diffusion-based image super-resolution in a single step. In: Proceedings of the IEEE/CVF Conference on Computer Vision and Pattern Recognition. pp. 25796--25805 (2024)

\bibitem{wang2022diffusion}
Wang, Z., Zheng, H., He, P., Chen, W., Zhou, M.: Diffusion-gan: Training gans with diffusion. arXiv preprint arXiv:2206.02262  (2022)

\bibitem{wang2004image}
Wang, Z., Bovik, A.C., Sheikh, H.R., Simoncelli, E.P.: Image quality assessment: from error visibility to structural similarity. IEEE transactions on image processing  \textbf{13}(4),  600--612 (2004)

\bibitem{wolters2023zooming}
Wolters, P., Bastani, F., Kembhavi, A.: Zooming out on zooming in: Advancing super-resolution for remote sensing. arXiv preprint arXiv:2311.18082  (2023)

\bibitem{xiao2024single}
Xiao, H., Wang, X., Wang, J., Cai, J.Y., Deng, J.H., Yan, J.K., Tang, Y.D.: Single image super-resolution with denoising diffusion gans. Scientific Reports  \textbf{14}(1), ~4272 (2024)

\bibitem{xiao2021tackling}
Xiao, Z., Kreis, K., Vahdat, A.: Tackling the generative learning trilemma with denoising diffusion gans. arXiv preprint arXiv:2112.07804  (2021)

\bibitem{xu2024ufogen}
Xu, Y., Zhao, Y., Xiao, Z., Hou, T.: Ufogen: You forward once large scale text-to-image generation via diffusion gans. In: Proceedings of the IEEE/CVF Conference on Computer Vision and Pattern Recognition. pp. 8196--8206 (2024)

\bibitem{yang2024structure}
Yang, L., Qian, H., Zhang, Z., Liu, J., Cui, B.: Structure-guided adversarial training of diffusion models. In: Proceedings of the IEEE/CVF Conference on Computer Vision and Pattern Recognition. pp. 7256--7266 (2024)

\bibitem{yang2023pixel}
Yang, T., Wu, R., Ren, P., Xie, X., Zhang, L.: Pixel-aware stable diffusion for realistic image super-resolution and personalized stylization. arXiv preprint arXiv:2308.14469  (2023)

\bibitem{yue2024resshift}
Yue, Z., Wang, J., Loy, C.C.: Resshift: Efficient diffusion model for image super-resolution by residual shifting. Advances in Neural Information Processing Systems  \textbf{36} (2024)

\bibitem{zeyde2012single}
Zeyde, R., Elad, M., Protter, M.: On single image scale-up using sparse-representations. In: Curves and Surfaces: 7th International Conference, Avignon, France, June 24-30, 2010, Revised Selected Papers 7. pp. 711--730. Springer (2012)

\bibitem{zhang2021designing}
Zhang, K., Liang, J., Van~Gool, L., Timofte, R.: Designing a practical degradation model for deep blind image super-resolution. In: Proceedings of the IEEE/CVF International Conference on Computer Vision. pp. 4791--4800 (2021)

\bibitem{zhang2018unreasonable}
Zhang, R., Isola, P., Efros, A.A., Shechtman, E., Wang, O.: The unreasonable effectiveness of deep features as a perceptual metric. In: Proceedings of the IEEE conference on computer vision and pattern recognition. pp. 586--595 (2018)

\end{thebibliography}
\bibliographystyle{springer_template/splncs04}




\end{document}